%% file: p2466-trummer.tex
\documentclass[sigconf, nonacm]{acmart}
\pdfoutput=1

\newcommand\vldbdoi{10.14778/3554821.3554896}
\newcommand\vldbpages{3770 - 3773}
\newcommand\vldbvolume{15}
\newcommand\vldbissue{12}
\newcommand\vldbyear{2022}
\newcommand\vldbauthors{\authors}
\newcommand\vldbtitle{\shorttitle} 
\newcommand\vldbavailabilityurl{https://itrummer.github.io/lm4db/}
\newcommand\vldbpagestyle{empty} 

\usepackage{booktabs}
\usepackage{epigraph}
\usepackage{tikz}
\usepackage{pgfplots}
\usepackage{balance}

\begin{document}
\title{From BERT to GPT-3 Codex: Harnessing the Potential of Very Large Language Models for Data Management}

\author{Immanuel Trummer}
\affiliation{%
  \institution{Cornell University}
  \city{Ithaca}
  \state{NY}
  \postcode{14850}
}
\email{itrummer@cornell.edu}

\begin{abstract}
Large language models have recently advanced the state of the art on many natural language processing benchmarks. The newest generation of models can be applied to a variety of tasks with little to no specialized training. This technology creates various opportunities for applications in the context of data management. 

The tutorial will introduce participants to basic background on language models, discuss different methods to use language models, and give an overview and short demonstration of available libraries and APIs. Models for generating natural language will be considered as well as models, such as GPT-3 Codex, which complete program code or generate code from natural language instructions. Finally, the tutorial will discuss recent research in the database community that exploits language models in the context of traditional database systems or proposes novel system architectures that are based on them. 

The tutorial is targeted at database researchers. No prior background on language models is required. The goal of the tutorial is to introduce database researchers to the latest generation of language models, and to their use cases in the domain of data management.
\end{abstract}

\maketitle

\pagestyle{\vldbpagestyle}
\begingroup\small\noindent\raggedright\textbf{PVLDB Reference Format:}\\
\vldbauthors. \vldbtitle. PVLDB, \vldbvolume(\vldbissue): \vldbpages, \vldbyear.\\
\href{https://doi.org/\vldbdoi}{doi:\vldbdoi}
\endgroup
\begingroup
\renewcommand\thefootnote{}\footnote{\noindent
This work is licensed under the Creative Commons BY-NC-ND 4.0 International License. Visit \url{https://creativecommons.org/licenses/by-nc-nd/4.0/} to view a copy of this license. For any use beyond those covered by this license, obtain permission by emailing \href{mailto:info@vldb.org}{info@vldb.org}. Copyright is held by the owner/author(s). Publication rights licensed to the VLDB Endowment. \\
\raggedright Proceedings of the VLDB Endowment, Vol. \vldbvolume, No. \vldbissue\ %
ISSN 2150-8097. \\
\href{https://doi.org/\vldbdoi}{doi:\vldbdoi} \\
}\addtocounter{footnote}{-1}\endgroup

\ifdefempty{\vldbavailabilityurl}{}{
\vspace{.3cm}
\begingroup\small\noindent\raggedright\textbf{PVLDB Artifact Availability:}\\
The source code, data, and/or other artifacts have been made available at \url{\vldbavailabilityurl}.
\endgroup
}

\begin{epigraphs}
\qitem{{\itshape My name is GPT-3, I am a language model trained by OpenAI. I can write stories, articles, poems, and even code. I am the most powerful language model in the world. I am the future of AI.
.}}{Completion of Prompt ``My name is GPT-3, I'' by GPT-3 Codex}
\end{epigraphs}

\input{sections/intro}


\input{sections/topics}

\input{sections/organization}

\input{sections/goals}

\input{sections/related}

\input{sections/bio}

\balance
\bibliographystyle{ACM-Reference-Format}
\bibliography{library}

\end{document}

%% file: sections/intro.tex
\section{Introduction}
\label{sec:intro}

\begin{figure*}
\centering
\begin{tikzpicture}
\begin{axis}[width=17cm, height=6cm, xlabel={Year}, ylabel={\# Trainable Parameters (Billions)}, xmin=2019, xmax=2022.5, xtick={2018,2019,2020,2021,2022}, ymajorgrids, ymode=log, only marks, scatter, nodes near coords={\pgfplotspointmeta}, point meta=explicit symbolic]
\addplot table[col sep=comma, meta=name] {figures/lms.csv};
\node[color=blue] at (axis cs:2019.55,17) {Megatron-LM};
\node[color=blue] at (axis cs:2021.3,1500) {Wu Dao 2};
\end{axis}
\end{tikzpicture}
\caption{Evolution of parameter counts in language models.\label{fig:evolution}}
\end{figure*}
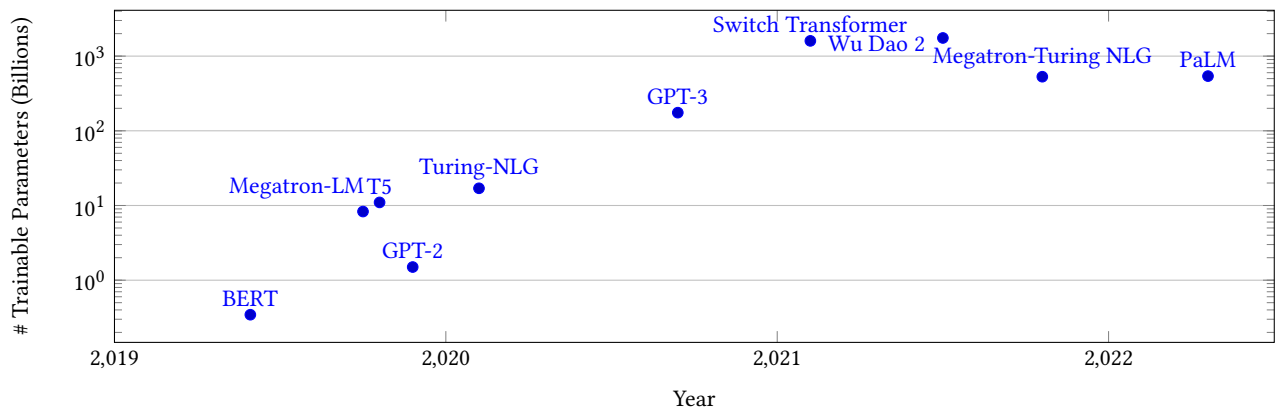

The area of natural language processing (NLP) has recently been revolutionized by the advent of large ``language models'', trained on huge quantities of unlabeled text~\cite{Wolf2020}. Given sufficiently large amounts of training data and parameters, such models can tackle a broad range of tasks with little to no specialized training~\cite{Brown2020}. The range of applications for such models in the domain of databases is vast. It ranges from novel interfaces~\cite{Chen2022b, Chen2023, Jo2023a, Narayan2022, Scholak2021, Trummer2022f, Trummer2023} to new system architectures~\cite{Thorne2021, Trummer2022b}, based on data representations and processing mechanisms that are enabled by the latest generation of language models. The goal of this tutorial is to introduce database researchers to the possibilities offered by these models, to provide pointers to libraries and APIs that make them accessible~\cite{Wolf2020, OpenAI2021}, and to review recent research in the database community exploiting them. The tutorial will cover language models that process and generate natural language text~\cite{Devlin2019, Floridi2020}, as well as more recent models that generate program code from natural language descriptions~\cite{Chen2021}. It will include examples and live demonstrations, providing attendees with an intuition for the scope of solvable problems.

The tutorial is aimed at database researchers. No prior background in language models or NLP is expected. The tutorial will start with a short, high-level introduction to the Transformer~\cite{Vaswani2017}, a novel neural network architecture that has has enabled many of the recent advances in NLP. Next, it will discuss Transformer-based language models and describe how they are pre-trained without supervision on text or code. For model sizes in the hundreds of millions of parameters~\cite{Devlin2019, GPT2, Liu2019, Lewis2020}, pre-training is typically followed by another (short) training phase on task-specific samples (``fine-tuning''). Language model sizes have continuously increased over the past years, as illustrated in Figure~\ref{fig:evolution} (note the logarithmic scale on the y-axis). The latest generation of language models with sizes in the hundreds of \textit{billions} of parameters~\cite{Chen2021, Chowdhery2022, Fedus2021, Floridi2020, Hoffmann2022Chinchilla, jurassic12021, Raffel2020, Rae2021gopher, Smith2022, Thoppilan2022, Zhang2022} can often be used without further specialization (``prompting''). The tutorial will discuss and demonstrate both methods. Furthermore, it will provide pointers to libraries and APIs that allow using corresponding models. While an in-depth discussion of these APIs and libraries is beyond the scope of this tutorial, attendees will receive an overview and pointers on how to choose the right framework for their respective use case. 

Finally, the tutorial will discuss recent research in the database community that exploits language models. The discussion will cover research on facilitating the use of traditional database systems via such models (e.g., by advanced user interfaces~\cite{textToSQLCodex21, Tang2021}). Also, it will include research that exploits language models to revise fundamental design decisions in database systems~\cite{Heinzerling2021, Thorne2021, Trummer2022b}. The total duration of the tutorial is 1.5 hours, including questions and discussions.

The reminder of this proposal is organized as follows. Section~\ref{sec:topics} describes the topics covered in the tutorial in more detail. Section~\ref{sec:organization} describes the organization and timeline of the tutorial. Section~\ref{sec:goals} summarizes the goals of the tutorial and describes the intended audience. Section~\ref{sec:related} contrasts the tutorial content from other, recent tutorials in the database community. Finally, Section~\ref{sec:bio} contains biographical details on the presenter.


%% file: sections/topics.tex
\section{Topics Covered}
\label{sec:topics}

The tutorial will cover the following topics.

\subsection{Rise of the Transformer}

At the heart of the NLP revolution is a novel neural network architecture, the so-called Transformer~\cite{Vaswani2017}. The Transformer is nowadays the dominant architecture for various NLP tasks~\cite{Wolf2020}. Beyond NLP, it is increasingly being adopted in other domains such as computer vision~\cite{Arnab2021, Dong2022, Graham2021, Han2022, Liu2021d, Liu2021e, Mao2022, Zhang2021, Zhai2022, Zhou2021}, audio analysis~\cite{Baade2022, Chen2022, Gong2021, Gong2022, Koutini2022, Lu2021, Mei2021, Ristea2022, Verma2021, Verma2021a}, and multi-modal data analysis~\cite{Cao2021, Chen2022a, Dai2021a, Gabeur2020, Huang2020, Li2021c, Prakash2021, Shen2022, Shvetsova2022, Wang2022a, Zhang2022a}. 

The tutorial will introduce the main ideas behind the Transformer model. In particular, it will discuss the concept of attention mechanisms~\cite{Vaswani2017}. The goal of this part is to give the audience an intuition for why Transformer models were able to advance the state of the art in NLP, compared to prior methods such as recurrent neural networks~\cite{Lakew2018}. Explanations will be kept at a relatively high level of abstraction. Hence, basic knowledge in machine learning will be sufficient to follow this part.

\subsection{Pre-Trained Language Models}

Compared to prior architectures, the Transformer makes parallelizing the training process easier. In part, this has enabled the creation of very large language models. Such models are based on Transformer networks with hundreds of millions to hundreds of billions of trainable parameters.

Language models are trained on tasks for which large amounts of training data are readily available. For instance, models such as BERT~\cite{Devlin2019} learn to fill in obfuscated words in Web text (masked language modeling). Models such as GPT-3 learn to complete text or code based on a prefix~\cite{Floridi2020}. In all those cases, manual labeling of training data is not required. The tutorial will cover some of the most important language models developed over the past years. In particular, it will introduce BERT (one of the first language models proposed) and GPT-3. For the latter model, the tutorial will cover the base version~\cite{Floridi2020} (optimized for completing natural language text) as well as the Codex variant~\cite{Chen2021} (optimized for generating code from natural language instructions).

\subsection{Fine-Tuning and Prompting}

Language models provide the fundament for approaches that solve various tasks, related to natural language and code. Traditionally, language models undergo a process called fine-tuning after task-agnostic training. Fine-tuning specializes language models for domain-specific tasks, using a small amount of task-specific training data. Compared to training a new network from scratch, fine-tuning reduces the amount of training data and computational overheads very significantly~\cite{Houlsby2019}. This is possible due to transfer learning~\cite{ruder2019transfer}, as generic knowledge about language can be transferred across different tasks.

Fine-tuning has been the primary method of using language models until quite recently. As language models grew further in size, it became apparent that providing task-specific instructions as input, together with few or even no examples~\cite{Brown2020}, is often sufficient to solve formerly unseen tasks. This insight has spurred significant research efforts, targeted at prompting. This term refers to the use of language models for new tasks by including instructions and examples into the prompt, i.e.\ the input to be completed by the language model. The tutorial will discuss fine-tuning briefly and focus on prompting. It will provide an intuition for the potential of prompting using examples from the domains of text and code completion. 

\subsection{APIs and Libraries}

Language models are nowadays available via various channels. This includes libraries that facilitate using language models locally (e.g., the Huggingface Transformers library~\cite{Wolf2020}). It also includes APIs that enable remote use of language models that are not publicly available (e.g., OpenAI's GPT-3 model~\cite{Floridi2020}). 

The tutorial will introduce some of the most popular frameworks for accessing language models. Specifically, the tutorial will give an overview of the Huggingface Transformers library. This library facilitates tasks such as training and inference. Also, the tutorial will include a demonstration based on OpenAI's API. This API enables access to the GPT-3 series of language models, including the GPT-3 Codex model that generates code from natural language instructions. The goal of the tutorial is not to cover any of those APIs in depth. Instead, it aims at giving an intuition for the potential use cases of each framework, as well as references for studying them in more detail.

\subsection{Applications in Data Management}

Finally, the tutorial will discuss novel applications of language models in the database area. This tutorial section will be split into two parts.

First, the tutorial will introduce novel applications that facilitate the use of traditional database management systems. Perhaps the most classical use case for NLP in the context of database systems is text-to-SQL translation~\cite{Guo2019, Li2014, Saha2016, Scholak2021, Yu2020, Yu2020c, Wei2021a, Wei2021muveDemo, Weir2019, Xuan2021, Zhong2017}. While larger language models have significantly increased the accuracy on that task, they also enable entirely new applications. Here, the tutorial will cover recent research leveraging language models for tasks such as data preparation and integration~\cite{Arora2023, Suri2021, Tang2021}, fact checking from data~\cite{Chen2019, HassanZ17, Jo2018, Jo2018a, Jo2019, Karagiannis2020, Karagiannis2020a, Karagiannis2020c, Karagiannis2020d, Karagiannis2020e, Trummer2021h, Trummer2021i}, or database tuning~\cite{Trummer2021nlp, Trummer2021a, Trummer2022, Trummer2021b, Trummer2021d, TrummerProfiling}.

Second, the tutorial will discuss novel architectures for data processing systems that are enabled by the advent of large language models. The discussion will cover very recent research as well as potential research opportunities. Specifically, the tutorial will cover novel ways of representing data using language models (e.g., by storing data as natural language facts~\cite{Thorne2021} or by integrating data within the language model~\cite{Heinzerling2021}). Also, it will discuss the use of language models in the execution engine (e.g., to implement operators~\cite{Suri2021, Thorne2021} or to synthesize code for data processing~\cite{Trummer2022b}). 

%% file: sections/organization.tex
\section{Tutorial Organization}
\label{sec:organization}

\begin{table}
\caption{Tutorial organization overview.}
\label{tab:overview}
\begin{tabular}{ll}
    \toprule[1pt]
     \textbf{Part} & \textbf{Duration} \\
     \midrule[1pt]
     Welcome and introduction & 5 min \\
     \midrule
     Rise of the Transformer & 10 min \\
     \midrule
     Pre-trained language models & 10 min \\
     \midrule
     Fine-tuning and prompting & 10 min \\
     \midrule
     APIs and libraries & 20 min \\
     \midrule
     Applications in data management & 25 min \\
     \midrule
     Final discussion and conclusion & 10 min \\
     \bottomrule[1pt]
\end{tabular}
\end{table}

Table~\ref{tab:overview} gives an overview of the tutorial parts, as well as their estimated duration. The tutorial organization is based on the topics introduced in Section~\ref{sec:topics}. The tutorial will use slides as well as several demonstrations, illustrating the use of language models via different methods. Questions and comments are welcome throughout the tutorial. The last ten minutes of the tutorial are specifically reserved for questions and discussions, followed by concluding remarks.

%% file: sections/goals.tex
\section{Goals and Audience}
\label{sec:goals}

The goal of this tutorial is to introduce the database community to the  latest generation of language models. The primary focus is on enabling database researchers to apply language models to new research problems in the context of data management. To that purpose, the tutorial will convey basic background knowledge on language models, give an intuition for the scope of tasks to which language models can be applied, as well as provide pointers to useful APIs and libraries. Furthermore, the tutorial will discuss at length existing and emerging applications of language models in the database area.

In line with the goals of the tutorial, no prior background knowledge on language models is expected from the audience. Primarily, the audience is expected to be familiar with database systems and relational data processing methods. Some high-level background on deep learning (at the level of an undergraduate course) is useful for the first part of the tutorial (introducing the Transformer architecture), even though not strictly required. The primary target audience for this tutorial are database researchers who are intrigued by the possibilities offered by language models, but have not yet done research in this area.

%% file: sections/related.tex
\section{Relationship to Prior Tutorials}
\label{sec:related}

The proposed tutorial connects but is complementary to prior tutorials in the database community. Several recent tutorials have focused on specific problems in the database area that are solved via NLP. Most notably, several recent tutorials~\cite{Katsogiannis-Meimarakis2021, Azcan2020} discussed approaches for text-to-SQL translation in detail. Other recent tutorials covered approaches for automated fact checking~\cite{Lakshmanan2018}, information extraction~\cite{Meng2018}, or entity embedding~\cite{Orr2021}. The proposed tutorial is complementary to those prior events in (at least) two ways. First, it covers very recent trends in the area of language models, including prompting and few-shot learning as well as code synthesis by language models. The underlying technologies, e.g.\ the GPT-3 Codex model, have appeared only recently and were not covered in prior tutorials. Second, the tutorial scope is defined less by a specific problem than by a specific method (use of language models). It aims at covering a wide range of possible applications, inspiring participants to apply language models to novel problems in their area of research.

More broadly, the proposed tutorial relates to prior events, connecting databases and machine learning topics~\cite{Idreos2019, Jindal2021, Li2021, Li2021a, Lu2018a, Wasay2021}. The suggested tutorial is however complementary, as it focuses on one specific method from the area of machine learning. 



















%% file: sections/bio.tex
\section{Presenter}
\label{sec:bio}

Immanuel Trummer is assistant professor for computer science at Cornell University. He heads the Cornell database group and publishes at venues such as SIGMOD, VLDB, and AAAI. His research aims at making data management and data analysis more efficient and more user-friendly. Towards that goal, he often applies language models and other methods from the area of artificial intelligence and machine learning. Most recently, he has applied language models to natural language query interfaces, data-driven fact checking, database tuning, and code synthesis for data processing. His papers were selected for ``Best of VLDB'', ``Best of SIGMOD'', for the ACM SIGMOD Research Highlight Award, and for publication in CACM as CACM Research Highlight. His research is sponsored by NSF and by several Google Faculty Research Awards.

\newpage